\documentstyle[aps]{revtex}
\begin{document}
\widetext
\draft

\title{Levinson's theorem and scattering phase shift contributions 
to the partition function of interacting gases in two dimensions}

\author{M. E. Portnoi\cite{byline} and I. Galbraith} 
\address{
Physics Department, Heriot-Watt University, Edinburgh EH14 4AS, 
United Kingdom} 

\date{Published in Phys. Rev. B 58, 3963 (1998)}

\maketitle

\begin{abstract}

We consider scattering state contributions to the partition function 
of a two-dimensional (2D) plasma in addition to the bound-state sum. 
A partition function continuity requirement is used to provide  
a statistical mechanical heuristic proof of Levinson's theorem 
in two dimensions. We show that a proper account of scattering 
eliminates singularities in thermodynamic properties of the nonideal 
2D gas caused by the emergence of additional bound states as the 
strength of an attractive potential is increased.
The bound-state contribution to the partition function of the 2D gas, 
with a weak short-range attraction between its particles, is found to 
vanish logarithmically as the binding energy decreases. 
A consistent treatment of bound and scattering states in a screened 
Coulomb potential allowed us to calculate the quantum-mechanical 
second virial coefficient of the dilute 2D electron-hole plasma 
and to establish the difference between the nearly ideal electron-hole 
gas in GaAs and the strongly correlated exciton/free-carrier plasma 
in wide-gap semiconductors such as ZnSe or GaN. 
\end{abstract}

\pacs{ 73.20.Dx, 71.35.-y, 05.30.-d, 03.65.Nk }

\section{INTRODUCTION}
\label{intro}

Two-dimensional (2D) systems play a central role in contemporary
condensed matter physics. 
Novel phenomena such as the quantum Hall effect \cite{QHE} 
observed when a 2D electron gas at low temperature is subjected 
to a strong magnetic field, 
as well as practical developments based on quasi-2D systems, 
e.g., high-mobility field-effect transistors \cite{AFS} 
or semiconductor quantum-well lasers \cite{Lasers}
have brought significant technological advances. 
Such devices are based on the quasiequilibrium response of the 
internal electron or electron-hole plasmas to an external stimulation. 
Hence it is essential to understand the fundamental quantum-statistical 
properties of such two-dimensional interacting plasmas at finite temperatures. 

One of the well-known differences between 2D and 3D nonrelativistic 
quantum mechanics is the presence of at least one bound state 
for any symmetric attractive potential in two dimensions. 
This bound state, with binding energy $~E_b$, brings a non-vanishing 
contribution of $~\exp(E_b/k_B T)~$ to the two-body part of the 
partition function even if the interaction is weak and the state 
is very shallow. 
For a dilute gas this contribution introduces a deviation from the 
ideal gas law that is larger than the correction 
due to the Fermi or Bose statistics of the particles. 
However, it is clear that 2D gases with a vanishing inter-particle interaction 
strength should be well described by free Fermi or Bose gas models. 
This contradiction will be resolved in the present paper.
A related question is what happens to the partition function when additional 
bound states appear with increasing strength of interaction between 
the particles. 
In the 3D case the answer to this question is based on a careful 
consideration of states in the continuum, which are modified by the 
interaction, i.e., taking scattering into account in the partition 
function calculation \cite{Larkin60,Rogers71}. 
The same approach should be applied in two dimensions, however, scattering 
theory in two dimensions is relatively undeveloped compared to the 3D case. 
For example, the relation between low-energy scattering and bound states, 
which has important consequences in the statistical mechanics, 
has been considered only very recently \cite{PG97,Lin97} in two dimensions. 
In the present paper the connection between scattering and 
the statistical mechanics of a 2D plasma is studied.

In the next section we introduce the relation of the two-dimensional 
scattering phase shift to the partition function and show that a proper 
consideration of the scattering states removes discontinuities in the 
partition function in an analogous fashion to the 3D case. 
In Sects.~\ref{short-range} and \ref{ehplasma} we explore this in more 
detail using an analytical model with an attractive square well potential 
before turning to a more realistic model of the statically screened 
electron-hole plasma, which is the main focus of this paper. 
We also compare in both cases the influence on the second virial 
coefficient of the exchange interaction and the screened direct interaction. 
Such considerations are crucial in understanding the nature of the 
strongly correlated electron-hole plasma in semiconductor quantum wells.
           
\section{PARTITION FUNCTIONS AND LEVINSON'S THEOREM}
\label{back}

The two-body interaction part of the partition function of 2D
interacting Boltzmann particles is given by 
\begin {eqnarray}
Z_{int}&~=~&\sum_{m,\nu} \exp(-E_{m,\nu}/k_B T)
\nonumber\\
&~+~&{1\over\pi} \int_0^\infty \sum_{m=-\infty}^{\infty} 
{d\delta_m(q) \over dq} \exp(-q^2 /q_T^2)~ dq ,
\label{eq1} 
\end {eqnarray}
where $~q_T^2=2\mu k_B T/\hbar^2~$,$~\mu~$ is the reduced mass, 
$~m~$ is the projection of the angular momentum onto the axis normal 
to the plane of 2D motion ($m=0, \pm 1, \pm 2, ~\ldots$), 
$~\delta_{m}(q)~$ is the 2D scattering phase shift \cite{SH67} dependent 
on the relative-motion momentum $~q$,
$~E_{m,\nu}~$ are the bound-state energies
(index $~\nu~$ enumerates bound states with given $~m$),
and the double sum in the first term ranges only over bound states. 
Eq.~(\ref{eq1}) is the 2D analogue of the Beth-Uhlenbeck formula and 
can be derived in the same fashion as in the 3D case \cite{BU37}. 

Often only the first term in Eq.~(\ref{eq1}) is considered when calculating 
an internal partition function, neglecting the phase shift term. 
For an attractive potential $~gU(\rho)~$, as $~g~$ decreases bound 
state energies increase towards the continuum. 
As such a state reaches zero energy, a partition function that
contains only the bound state sum will be discontinuous.
These unphysical singularities would extend to all the thermal 
properties, such as pressure and specific heat.  
Integrating by parts we can rewrite Eq.~(\ref{eq1}) as  
\begin {eqnarray}
Z_{int}&~=~&\left\{\sum_{m,\nu} \exp(-E_{m,\nu}/k_B T)
-\sum_{m=-\infty}^{\infty} \delta_m(0)/\pi\right\}
\nonumber \\
&~+~&{2\over{\pi q_T^2}} 
\int_0^\infty \left(\sum_{m=-\infty}^{\infty} \delta_m(q)\right) 
\exp(-q^2/q_T^2) q dq~.
\label{eq2} 
\end {eqnarray}
For nonzero temperature the integral term in the right-hand side 
of Eq.~(\ref{eq2}) is a smooth function of the interaction strength $~g$. 
If the phase shifts satisfy the condition:
\begin {equation} 
\lim_{q \rightarrow 0} \delta_m(q)~=~\nu_m\pi~, 
\label{eq3} 
\end {equation}
where $~\nu_m~$ is the number of bound states with given $~m~$,
the zero-energy part of the phase shift integral in Eq.~(\ref{eq1})
exactly cancels the zero-energy part of the bound state sum, removing 
the discontinuity in $~Z_{int}~$ as a function of the interaction strength.
This  cancellation is similar to the well known behaviour in 3D where 
the partition function discontinuities are removed \cite{Rogers71} 
with the help of Levinson's theorem \cite{Levinson}.
Equation (\ref{eq3}) constitutes the 2D statement of Levinson's theorem. 

As a central theorem of scattering theory \cite{Newton}, 
Levinson's theorem  has been discussed for Dirac 
particles, multichannel scattering, multi-particle single-channel scattering, 
one-dimensional scattering systems, impurities in Aharonov-Bohm rings,
systems with non-uniform effective mass, 
and even for time-periodic potentials \cite{NewLevinson}. 
However, its applicability to the 2D scattering problem has been 
considered only recently. 
In Ref.~\cite{PG97} the 2D statement of Levinson's theorem,  
Eq.~(\ref{eq3}), was proposed and verified empirically, 
while in Ref.~\cite{Lin97} this theorem was more rigorously established 
for cutoff potentials using the Green-function method. 
The above arguments, based on the partition function continuity requirement, 
provide an additional statistical mechanical justification of Levinson's 
theorem in 2D. 

The two-body interaction part of the partition function can be used  
to calculate the second virial coefficient, $~B(T)$, 
which characterizes the first correction to the ideal gas 
law in the low-density expansion of pressure 
\begin {equation} 
P~=~n k_B T~(1~+~B n~+~\ldots)~. 
\label{eq4} 
\end {equation} 
$B~$ is positive for repulsive potentials, 
causing an increase in the pressure over its ideal-gas value, 
and negative for attractive potentials, causing a decrease in the pressure.
A calculation of the second virial coefficient is meaningful in the dilute 
gas regime, where the mean inter-particle spacing $~n^{-1/2}~$ is larger 
than the thermal wavelength $~\lambda=(2 \pi \hbar^2 / M k_B T)^{1/2}~$ 
and the higher order terms in Eq.~(\ref{eq4}) are negligible. 
For a free, Bose or Fermi gas in two dimensions, 
$~B(T)=\mp\lambda^2/4~$ \cite{SS73D75}, 
the plus sign applying to the Fermi case, 
as the Pauli principle introduces an effective repulsion between fermions, 
thereby increasing the pressure. 
There is extensive ongoing research in the statistical mechanics of 
anyons, 2D particles obeying fractional statistics \cite{Anyons}, 
and the second virial coefficient of a free anyon gas lies in between 
the bosonic and fermionic value \cite{Arovas85}. 
In this paper we focus on the relative importance  of interaction-induced 
bound and scattering states on the second virial coefficient.

The 3D analysis of the second virial coefficient \cite{Kilpatrick} 
is easily reformulated for a 2D interacting gas \cite{SS73D75}. 
For a system of identical particles with spin $s$ the second virial 
coefficient is
\begin {equation} 
B^{(s)}(T)~=~{\lambda^2 \over 2s+1} \left(\mp {1 \over 4}
~-~{2 Z^{(s)}_{int} \over 2s+1} \right)~,
\label{eq5}
\end {equation}
where the upper sign is for bosons and the lower sign for fermions.
The exclusion principle modifies the sum over $~m~$ in Eq.~(\ref{eq2}) 
depending on the angular momentum parity, 
and the partition function (using Eq.~(\ref{eq3})) is
\begin {eqnarray}
Z^{(s)}_{int}&~=~& (2s+1) \sum_{m,\nu}
\left(s+{1\pm(-1)^m \over 2} \right) [\exp(-E_{m,\nu}/k_B T)-1]
\nonumber \\
&&
\nonumber\\
&~+~&(2s+1){2\over{\pi q_T^2}} \int_0^\infty q dq 
\left[\sum_{m=-\infty}^{\infty} 
\left(s+{1\pm(-1)^m \over 2}\right) \delta_m(q) \right] 
\exp(-q^2/q_T^2)~.
\label{eq6} 
\end {eqnarray}

The electron-hole plasma  constitutes a mixture of two components and  
for a binary mixture of components $~C~$ and $~D~$ having second virial 
coefficients $~B^{(s)}_C~$ and $~B^{(s')}_D~$ and densities $~n_C$ and
$~n_D$, respectively, the second virial coefficient is \cite{SS73D75}
\begin {equation} 
B(T)~=~\left({n_C \over n}\right)^2 B^{(s)}_C~+~
2{n_C n_D\over n^2} B_{CD}~+~\left({n_D \over n}\right)^2 B^{(s')}_D~.
\label{eq7}
\end {equation}
In Eq.~(\ref{eq7}) $~n=n_C+n_D~$ and 
\begin {equation}
B_{CD}=-\lambda^2_\mu Z_{int}
\label{eq8}
\end {equation}
where
\begin {equation}
\lambda^2_\mu~=~{2\pi \hbar^2 \over 2 \mu k_B T}~,
~~\mu~=~{M_C M_D \over M_C+M_D}~, 
\label{eq9}
\end {equation}
and $~Z_{int}~$ is given by Eq.~(\ref{eq1}) or Eqs.~(\ref{eq2})
and (\ref{eq3}) with the properly chosen reduced mass $~\mu$.

\section{BOLTZMANN GAS WITH SHORT-RANGE ATTRACTION}
\label{short-range} 

Our first example, the Boltzmann gas with weak short-range attraction, 
is chosen to elucidate how the second virial coefficient at given 
temperature vanishes when the binding energy $~E_b~$ decreases, 
even though the bound-state part of the internal partition function 
$~Z_{bound}~=~\exp(E_b/k_B T)~$ approaches unity rather than vanishes. 
To trace the precise nature of the cancellation of $~Z_{bound}~$ it 
is convenient to use Eq.~(\ref{eq1}) for $~Z_{int}~$ without 
the application of Levinson's theorem.

Let us assume that the 2D particles interact via an attractive 
square-well potential of radius $a$ and depth $~V_0$. 
This simple model allows analytical treatment which provides 
insight into the generic behaviour of a gas of attracting particles. 
To evaluate the partition function we need to analyse both the bound and
scattering states in this potential.
The binding energies  for any value of angular 
momentum $~m~$ can be easily found by matching the logarithmic 
derivative of the radial wave function at $~\rho=a~$:
\begin{equation}
{\sqrt{\kappa_0^2-\kappa^2} 
J_{|m|+1}\left(a\sqrt{\kappa_0^2-\kappa^2}\right)
\over J_{|m|}\left(a\sqrt{\kappa_0^2-\kappa^2} \right)}~=~
{\kappa K_{|m|+1}(\kappa a) \over K_{|m|}(\kappa a)}~,
\label{eq10}
\end{equation}
where $~\kappa_0^2=2 \mu V_0/\hbar^2~$, 
$~\kappa^2=2 \mu E_b/\hbar^2~$, 
$~J_m(x)~$ is the Bessel function of the first kind and $~K_m(x)~$ is the 
modified Bessel function of the second kind.
Note that  for $~\kappa_0 a < 2.4~$(i.e. smaller than 
the first root of $~J_0(x)~$) there is only one bound state 
(having  $~m=0$) and
for $~\kappa_0 a < 1~$ this state is very shallow,  e.g.
for $~\kappa_0 a = 1~$, $~E_b/V_0 \approx 0.04~$, for  
$~\kappa_0 a = 0.5~$, $~E_b/V_0 \approx 2 \times 10^{-7}$. 
For a shallow $~m=0~$ state the transcendental equation for binding
energies Eq.~(\ref{eq10}) reduces to
\begin{equation}
{\kappa_0 a J_1(\kappa_0 a) \over J_0(\kappa_0 a)}~=~
{\kappa a K_1(\kappa a) \over K_0(\kappa a)}~=~
-{1 \over \ln(c \kappa a)},
\label{eq11}
\end{equation}
where $~c=\exp(\gamma)/2~$ 
($~\gamma \approx 0.5772157 \dots~$ is Euler's constant).
 
For the unbound states with positive energy of the relative motion,
$~E=\hbar^2 q^2/2\mu~$, scattering phase shifts can be found
in a similar fashion. 
For small values of the momentum, $~qa \ll 1~$,
all phase shifts for $~m \neq 0~$ are small compared to $~\delta_0~$
($s$-wave scattering) \cite{LandauQM}. The tangent of the $s$-wave 
scattering phase shift for $~q \ll \kappa_0~$ (i.e. $~E \ll V_0$) 
is given by \cite{Portnoi88}:  
\begin{equation}
\tan {\delta_0}={\pi/2 \over \ln(cqa)+
{J_0(\kappa_0 a) \over \kappa_0 a J_1(\kappa_0 a)}}~.
\label{eq12}
\end{equation}
Substituting $~\kappa_0 a J_1(\kappa_0 a)/J_0(\kappa_0 a)~$ 
from Eq.~(\ref{eq11}) into Eq.~(\ref{eq12}) we get
\begin{equation}
\tan {\delta_0}={\pi \over \ln(E/E_b)}.
\label{eq13}
\end{equation}
Note that this expression does not contain parameters of the potential 
$~V_0~$ and $~a~$ explicitly, and it is valid for an arbitrary potential 
well with a shallow $m=0$ level \cite{xsect}. 

Since the integrand in the partition function, Eq.~(\ref{eq1}), 
contains an exponential factor $~\exp(-q^2/q_T^2)~$, the wavevectors 
$~q~$ which are larger than the thermal wave vector $~q_T~$ give
negligible contribution to the value of the integral. Therefore for 
the short-range interaction or for low temperature, satisfying 
condition $~k_B T \ll \hbar^2 / 2 \mu a^2~$, the scattering phase 
shifts need only be considered  for $~q \ll 1/a~$. Then all 
the terms in the phase shift sum in Eq.~(\ref{eq1}) can be neglected 
except for the term with $~m=0$. 
Finding the derivative of $~\delta_0~$ from Eq.~(\ref{eq13}) we 
obtain for the two-body interaction part of the partition function
\begin{equation}
Z_{int}~=~\exp(E_b/k_B T)~-~\int_0^\infty 
{\exp(-E/k_B T) \over \pi^2+\ln^2(E/E_b)} {dE \over E}~.
\label{eq15}
\end{equation}
The integral in Eq.~(\ref{eq15}) is the
Ramanujan integral \cite{Hardy} which  can 
be rewritten as \cite{Hardy,Ryzhik}
\begin{equation} 
\int_0^\infty {e^{-xt} \over {\pi^2+(\ln t)^2} } {dt \over t} ~=~
e^x~-~\nu(x),
\label{eq16}
\end{equation}
where
\begin{equation}
\nu(x)~=~\int_0^\infty {x^t \over \Gamma(t+1)} dt~,
\label{eq17}
\end{equation}
with $~x=E_b/k_B T~$. 
Thus, the partition function  acquires a very simple form 
\begin{equation}
Z_{int}~=~\nu(E_b/k_B T)~.
\label{eq18}
\end{equation}
A similar result has been obtained recently for contact-interacting 
particles \cite{Anyons}.

To consider the small $~x~$ asymptotic of the function $~\nu(x)~$ 
it is convenient to expand the integral 
Eq.~(\ref{eq17}) in descending powers of $~\ln(1/x)$:
\begin{equation} 
\nu(x)~=~{1 \over \ln(1/x)}~+~{\gamma \over \ln^2(1/x)}~+~
O\left([\ln(1/x)]^{-3}\right)~.
\label{eq19}
\end{equation} 
From Eq.~(\ref{eq19}) one can see that $~Z_{int}~$ and hence the 
second virial coefficient $~B=-\lambda^2 Z_{int}~$ both vanish
when $~E_b/k_B T \rightarrow 0~$, although one bound state always exists. 
So the lowest-order density correction to the 2D ideal gas law 
vanishes only slowly as $~1/\ln(k_B T/E_b)~$ as the binding energy
is reduced.
Note that, when the potential supports several bound states, 
the contribution of any shallow bound state with $~m=0~$ is cancelled 
by the scattering phase shift integral in the ``logarithmic'' manner 
described above. 
For $~m \neq 0~$ the cancellation has a power-law dependence 
in $~E_b/k_B T~$ \cite{PGunpubl}. 
This implies that higher-order Levinson's theorems responsible for 
continuity of the partition function derivatives \cite{Bolle} 
are different for $~m=0~$ and $m \neq 0$, 
whereas the zeroth-order Levinson's theorem in two dimensions has 
the same form,  Eq.~(\ref{eq3}),  for all $~m$.

For extremely weak interaction potential, such that
$~(\kappa_0 a)^2 \ll 1/\ln(\hbar^2/ 2 \mu a^2 k_B T)$, 
from Eqs.~(\ref{eq11}) and (\ref{eq19}) it follows that 
$~Z_{int}\approx V_0\mu a^2/2 \hbar^2$, which coincides with 
the perturbation theory result.
In the other limit for large $~x~$ values ($E_b/k_B T \gg 1$), 
$~\nu(x) \rightarrow e^x~$ \cite{HTF}, 
therefore the exponential dependence of the partition 
function on the binding energy is recovered. 

\begin{figure}
\begin{center}
\includegraphics{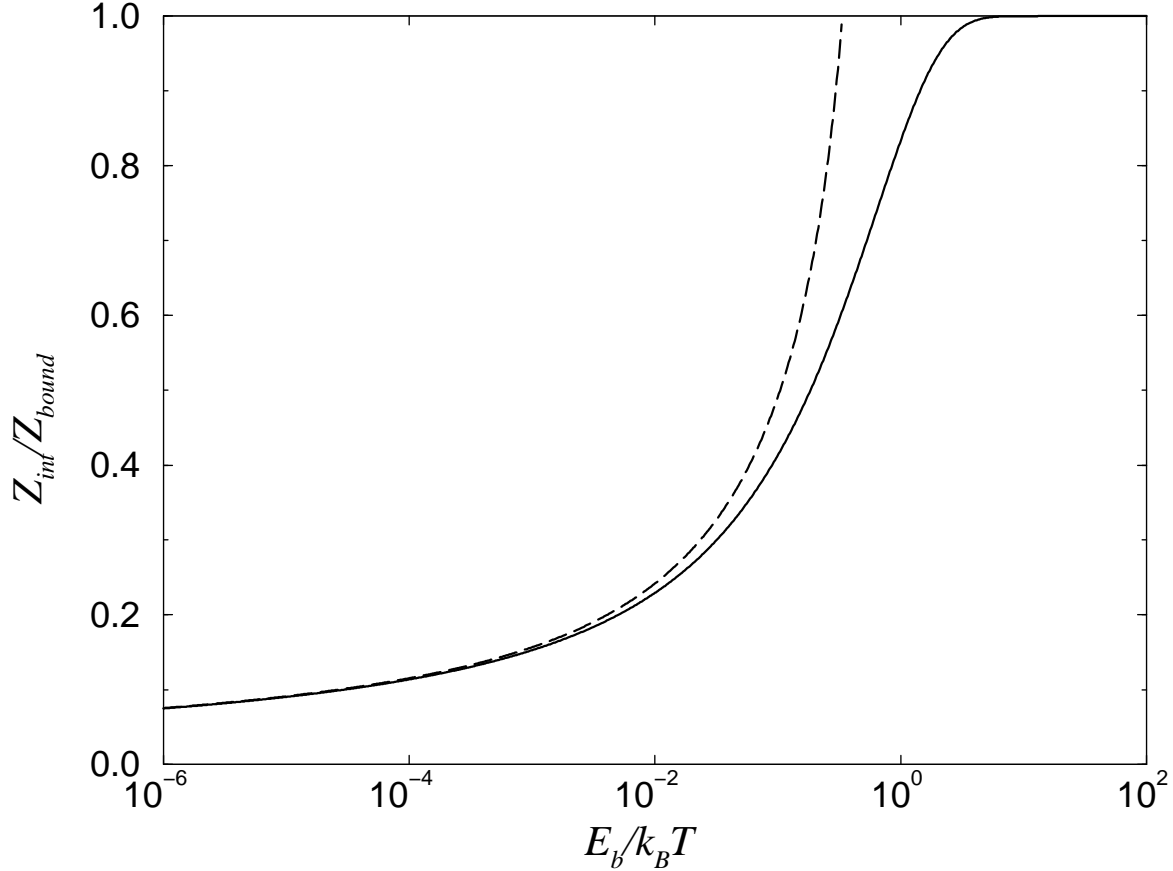}
\end{center}
\vskip 16truecm
\caption{
Two-dimensional Boltzmann gas with a short-range weak attraction: 
the ratio of the two-body partition function to its bound-state part,  
$~Z_{int}/Z_{bound}=\nu(E_b/k_B T)/\exp(E_b/k_B T)$, is
plotted versus $~E_b/k_B T$. 
Dashed line: the function $~\nu(x)~$ is approximated by 
$~1/\ln(1/x)+\gamma/\ln^2(1/x)$.}
\label{fig1}
\end{figure}

In Fig.~\ref{fig1} we plot the ratio of the total partition function 
$~Z_{int}$ to its bound-state part $~Z_{bound}=\exp(E_b/k_B T)~$  
as a function of $~E_b/k_B T$.
We do this for both the full expression, Eq.~(\ref{eq17}), and the 
first two terms in the asymptotic expansion, Eq.~(\ref{eq19}). 
One can see that the asymptotic expression  (dashed line in 
Fig.~\ref{fig1}) is accurate only for very small values of $~E_b/k_B T$. 
Over a wide range of $~E_b/k_B T~$ both scattering and bound-state 
terms are important, e.g. when $~E_b=k_B T~$ the scattering term 
produces a 20\% correction to $~Z_{int}$. 
When $~E_b/k_B T>3~$ the bound-state contribution dominates completely.

It is instructive to compare the contributions to the second virial 
coefficient of the direct and the exchange interactions.
For example for a gas of spinless bosons the second virial 
coefficient is $~B=-\lambda^2(1/4+2Z_{int})$. For $~Z_{int}~$ 
 given by Eq.~(\ref{eq18}), 
the direct interaction term ($2 Z_{int}$) is smaller than the exchange 
term ($1/4$) for 
  small binding energies.
Thus, as required for vanishing binding energy, the second virial 
coefficient is governed by the statistics of particles, 
despite the existence of a single bound state.

\section{ELECTRON-HOLE PLASMA}
\label{ehplasma}
In this section we study a more realistic  model of an interacting gas 
than the one considered in Sect.~\ref{short-range}.
We consider a mixture of the two types of Fermi particles, positively 
charged holes (h) and negatively charged electrons (e).
This model is important for understanding the thermodynamic and optical 
 properties 
of the electron-hole plasma in semiconductor quantum wells.
In the low-temperature, low-density limit most electron-hole pairs 
are bound into excitons. 
This limit has been studied extensively because of the recent 
proposals of exciton condensates in this system \cite{ZLHR95,FT97}. 
The properties of the degenerate 2D electron-hole plasma 
are also well known \cite{SCM89}. 
We consider the case when the temperature is comparable to the exciton 
binding energy so the occupation of continuum states (and therefore 
screening) is significant, although the carrier density is low enough to 
neglect the terms higher than $~Bn~$ in the virial expansion. 
The temperature and density conditions under investigation are close to 
those in the regime where excitonic gain in wide-gap semiconductors 
is anticipated \cite{Ian96}.
 
We assume for simplicity equal masses $~M_e=M_h=M~$ and 
spins $~s=s'=1/2~$ for both species. 
Then $~Z_{hh}=Z_{ee}~$ and the second virial coefficient 
for the mixture Eq.~(\ref{eq7}) acquires the form: 
\begin{equation} 
B~=~{\lambda^2 \over 4} \left({1 \over 4}~-~2Z_{eh}~-~Z_{ee}\right).
\label{eq20}
\end{equation}
Here $~\lambda=\lambda_{\mu}=(2 \pi \hbar^2 / M k_B T)^{1/2}~$
and the charge neutrality condition $~n_e=n_h$ is taken into account.

The screened Coulomb attraction between electrons and holes is modeled 
using the Fourier representation of the interaction potential:
\begin{equation}
V_q~=~-2 {2\pi \over q+q_s},
\label{eq21}
\end{equation}
where $~q_s~$ is the screening wave vector.
Hereafter we employ 3D excitonic Rydberg units 
where length and energy are scaled, respectively, 
by the effective Bohr radius $~a^*~$ and Rydberg Ry$^*$. 
For electron-electron and hole-hole repulsion the same potential 
with the opposite sign is used.
Equation (\ref{eq21}) is the well-known Thomas-Fermi expression for 
the Coulomb potential statically screened by a 2D electron gas.
Being the long-wavelength static limit of the random phase approximation,
Eq.~(\ref{eq21}) is a simple model for the screened Coulomb potential 
in two dimensions. 
Nevertheless, this expression reflects the fact that the statically 
screened potential in two dimensions decreases at large distances 
slower than in the 3D case (as a power law rather than exponentially). 
Despite numerous more realistic corrections \cite{AFS,screen97} 
Eq.~(\ref{eq21}) remains the most widely used approximation for the 
2D screening \cite{HKbook,WKB97}. 
This potential has been known for more than three decades \cite{SH67} 
but some of its unusual properties were only discovered recently, 
e.g. the existence of a remarkably simple relation between $~q_s~$ 
and the number of bound states. 
Namely, with decreasing screening, new bound states appear at the critical 
integer values of the screening length given by the simple formula\cite{PG97}
\begin{equation}
\left(1 \over q_s\right)_c~=~ 
{(2|m|+\nu-1)(2|m|+\nu) \over 2}~,~~~~\nu=1,2,~\ldots~,
\label{eq22}
\end{equation}
where $~m~$ is the angular momentum and $~(\nu-1)~$ indicates the 
number of nonzero nodes of the radial wave function.
Equation (\ref{eq22}) shows that several bound states corresponding 
to the given value of $~(2|m|+\nu)~$ appear simultaneously. 
This degeneracy is different from the degeneracy for the bound 
states of the unscreened 2D exciton (or hydrogen atom), for which 
the states with the same value of $~(|m|+\nu)~$ are degenerate 
\cite{HKbook,Flugge}. The hidden symmetry that underlines this 
new degeneracy has not been fully understood yet, and Eq.~(\ref{eq22}) 
still lacks a rigorous analytical derivation.

We also consider the low-density (nondegenerate) limit, when there is no 
Pauli blocking and the self-energy correction \cite{ZS85} to the 
Beth-Uhlenbeck formula can be neglected and Eqs.~(\ref{eq2},\ref{eq3}) 
and Eq.~(\ref{eq6}) can be used for $~Z_{eh}~$ and $~Z_{ee}~$ respectively. 
The shortcomings of this model for the quantitative description of a real 
system of photoexcited electrons and holes in semiconductor quantum wells 
are self-evident; however, it does provide a tractable model 
containing all the salient features of the  system. 

To find the second virial coefficient given by Eq.~(\ref{eq20}) 
one must calculate the binding energies and scattering 
phase shifts entering the partition functions $~Z_{eh}~$ and $~Z_{ee}~$.
We use for this purpose the 2D modification of the variable-phase 
method \cite{Cal67} known from scattering theory.   
In this method the scattering phase shift and the function 
defining bound-state energies can be obtained as a large 
distance limit of the phase function, which satisfies the 
first-order, nonlinear Riccati equation originating from the radial 
Schr\"{o}dinger equation. The variable-phase method application to 
scattering and bound states in the screened Coulomb potential 
(\ref{eq21}) is described in detail in Ref.~\cite{PG97}. The 
method is especially effective for calculation of shallow-state
binding energies and low-energy scattering phase shifts. 

\begin{figure}
\begin{center}
\includegraphics{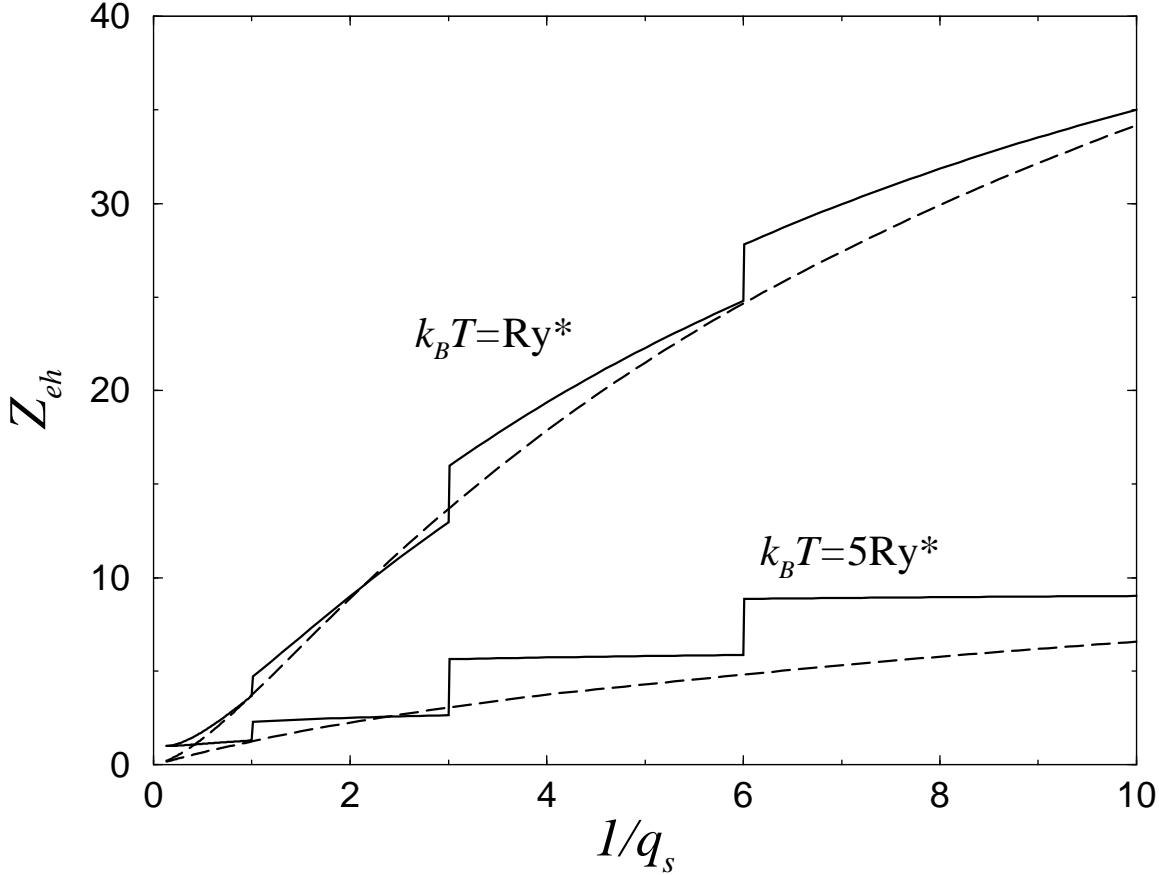}
\end{center}
\vskip 16truecm
\caption{
The electron-hole part of the partition function, $~Z_{eh}~$ 
versus the screening length $~1/q_s~$ for two values of 
$~k_B T/{\rm Ry}$. Solid lines show the bound state contributions 
$~Z_{bound}~$ only. Dashed lines: $~Z_{bound}+Z_{scatt}$.}
\label{fig2}
\end{figure}

Figure \ref{fig2} shows the results from the calculation of the 
electron-hole part of the partition function, $~Z_{eh}$, which contains 
both the bound state sum and the scattering phase shift integral.
In this figure $~Z_{eh}~$ is plotted as a function of the inverse 
screening wave number $~1/q_s$. 
For the  electron-hole plasma $~q_s~$ is a function of 
carrier density \cite{HKbook}, and in the purely 2D case 
$~q_s \propto (n_e+n_h)~$ for low densities and is 
independent of density, $~q_s \rightarrow 8/a^*~$ for $~M_e=M_h$, in 
the degenerate limit. Here we treat $~q_s~$ as 
a parameter characterizing the strength of the screened interaction 
potential Eq.~(\ref{eq22}). To emphasize the role of scattering 
we show on the same plot the bound-state sum, 
$~Z_{bound}=\sum_{m,\nu} \exp(-E_{m,\nu}/k_B T)~$, 
which exhibits jumps whenever new bound states appear 
[i.e. when $~q_s~$ satisfies Eq.~(\ref{eq22})]. These jumps become 
higher with increasing screening length $~1/q_s~$ since 
several bound states appear simultaneously. 
As can be seen the additional scattering phase shift contributions 
completely remove these jumps. 
The partition function is plotted for two values of the ratio 
of $~k_B T~$ to the excitonic Rydberg,
$~k_B T=1~{\rm Ry}^*$ and $~k_B T=5~{\rm Ry}^*$, which roughly 
correspond to ZnSe (or GaN) and GaAs at room temperature.  
One can see that for high temperature (or low binding energy) the 
bound-state contributions to $~Z_{eh}~$ are suppressed by the scattering 
phase shift integral more strongly than in the $~k_B T=1~{\rm Ry}^*~$ case.

\begin{figure}
\begin{center}
\includegraphics{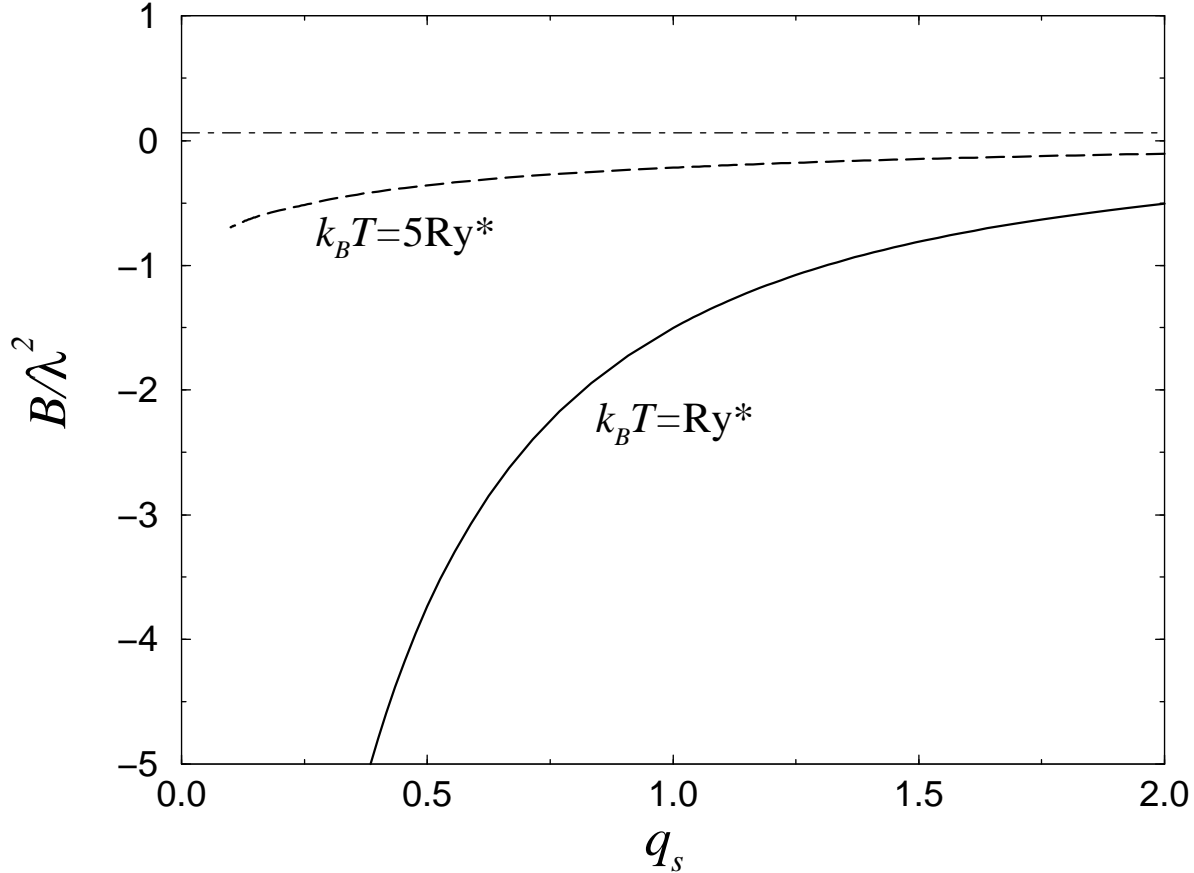}
\end{center}
\vskip 16truecm
\caption{
The second virial coefficient of the dilute electron-hole 
plasma $~B/\lambda^2~$ as a function of the screening 
wavenumber $~q_s~$. Solid line: $~k_B T=1~{\rm Ry}^*$;
dashed line $~k_B T=5~{\rm Ry}^*$; dot-dashed line: non-interacting 
dilute electron-hole plasma, $~B/\lambda^2=1/16$.}
\label{fig3}
\end{figure}

In Fig.~\ref{fig3} the second virial coefficient $~B~$ 
(scaled by $~\lambda^2$) is plotted versus the screening wave number 
$~q_s~$ for two different values of $~k_B T$. 
Equation (\ref{eq20}) is used for the calculation of $~B$, and the 
repulsion ($Z_{ee}$) term partially compensates the $~Z_{eh}~$ term. 
This compensation is especially significant in the high-temperature case
$~k_B T=5{\rm Ry}^*$, in which the 2D electron-hole plasma behaves much 
like an ideal gas over a wide range of screening wave vectors. 
For $~k_B T=1~{\rm Ry}^*~$ the electron-hole attraction term dominates 
and the plasma is strongly correlated for all values of $~q_s$. 
In this case a small statistical repulsion ($B/\lambda^2=1/16$, 
horizontal line in Fig.~\ref{fig3}),
which is due to the fermionic nature of electrons and holes, can 
be completely neglected. Thus, at room temperature the electron-hole  
plasma in GaAs-based quantum wells can be treated as an ideal gas, 
whereas in wide-gap semiconductors (e.g., ZnSe or GaN) 
due to the high value of Ry$^*$ the 2D electron-hole plasma is 
strongly correlated and excitonic effects are important for its 
thermodynamic properties.

\section{CONCLUSION}
\label{concl}

In this paper  we show that a proper 
account of scattering eliminates discontinuities in thermodynamic 
properties of the nonideal 2D gas whenever extra bound states appear 
with small increase of the strength of an attractive potential. 
This treatment provides a heuristic proof
of Levinson's theorem in two dimensions.

We trace the way in which the bound-state contribution to the partition 
function of the 2D gas, with a weak short-range attraction between its 
particles, vanishes when the binding energy decreases. 
A weak $~1/\ln(k_B T/E_b)~$ binding energy dependence 
of the second virial coefficient 
of such a gas is found for $~E_b/k_B T \rightarrow 0$.   

A consistent treatment of bound and scattering states in a screened 
Coulomb potential allows us to calculate the quantum-mechanical second 
virial coefficient of the dilute 2D electron-hole plasma and to establish 
the difference between the nearly ideal electron-hole gas in GaAs and the 
strongly correlated exciton/free-carrier plasma in wide-gap semiconductors.

The 2D electron-hole plasma was considered in the low-density nondegenerate 
limit only. Transition to the strongly degenerate Fermi limit and related 
questions of Pauli blocking and self-energy corrections to the 
Beth-Uhlenbeck formula in 2D remain the subject of further research.
  
\acknowledgments

This work was supported by the U.K. EPSRC and the Royal Society.


\begin{references}
\bibitem{xsect} It is interesting to note that the expression for the 
scattering cross-section for resonant $s$-wave scattering in two dimensions,
$~\sigma=(4/q) \sin^2{\delta_0}=(4\pi^2/q)/[\pi^2+\ln^2(E/E_b)]$, 
despite being quite different from its 3D equivalent \cite{LandauQM}, 
nevertheless also has a pole at $~E=-E_b$.
\end{references}
\end{document}